\documentclass[%
 reprint,
 amsmath,amssymb,
 aps,
prl,
superscriptaddress
]{revtex4-1}

\usepackage[dvipdfmx]{graphicx}
\usepackage{graphicx}
\usepackage{dcolumn}
\usepackage{bm}
\usepackage{color}
\usepackage{csquotes}

\begin{document}

\preprint{APS/123-QED}

\title{Structural transformations in tetravalent nematic shells induced by a magnetic field}

\author{Yoko Ishii}
  \affiliation{Department of Physics, Graduate School of Science, Kyoto University, Oiwake-cho, Kitashirakawa, Sakyo-ku, Kyoto, 606-8562, Japan\\}
\author{Ye Zhou}
 \affiliation{Pritzker School of Molecular Engineering, University of Chicago, Chicago, Illinois 60637, USA\\}
\author{Kunyun He}
\affiliation{UMR No.7083, CNRS, Gulliver, ESPCI Paris, PSL Research University, 10 Rue Vauquelin, 75005 Paris, France\\}
\author{Yoichi Takanishi} 
 \affiliation{Department of Physics, Graduate School of Science, Kyoto University, Oiwake-cho, Kitashirakawa, Sakyo-ku, Kyoto, 606-8562, Japan\\}
\author{Jun Yamamoto} 
 \affiliation{Department of Physics, Graduate School of Science, Kyoto University, Oiwake-cho, Kitashirakawa, Sakyo-ku, Kyoto, 606-8562, Japan\\}
\author{Juan de Pablo}
\email{depablo@uchicago.edu}
 \affiliation{Pritzker School of Molecular Engineering, University of Chicago, Chicago, Illinois 60637, USA\\}
\author{Teresa Lopez-Leon}
 \email{teresa.lopez-leon@espci.fr}
\affiliation{UMR No.7083, CNRS, Gulliver, ESPCI Paris, PSL Research University, 10 Rue Vauquelin, 75005 Paris, France\\}

\date{\today}

\begin{abstract}

The role of applied fields on the structure of liquid crystals confined to shell geometries has been studied in past theoretical work, providing strategies to produce liquid crystal shells with controlled defect structure or valence. However, the predictions of such studies have not been experimentally explored yet.  In this work, we study the structural transformations undergone by tetravalent nematic liquid crystal shells under a strong uniform magnetic field, using both experiments and simulations. We consider two different cases in terms of shell geometry and initial defect symmetry: i) homogeneous shells with four $s = +1/2$ defects in a tetrahedral arrangement, and ii) inhomogeneous shells with four $s = +1/2$ defects localized in their thinner parts. Consistently with previous theoretical results, we observe that the initial defect structure evolves into a bipolar one, in a process where the defects migrate towards the poles. Interestingly, we find that the defect trajectories and dynamics are controlled by curvature walls that connect the defects by pairs. Based on the angle between $\mathbf{B}_{\mathrm{s}}$, the local projection of the magnetic field on the shell surface, and  $\mathbf{n}_{+\frac{1}{2}}$, a vector describing the defect orientations, we are able to predict the nature and shape of those inversion walls, and therefore, the trajectory and dynamics of the defects. This rule, based on symmetry arguments, is consistent with both experiments and simulations and applies for shells that are either homogeneous or inhomogeneous in thickness. By modifying the angle between $\mathbf{B}_{\mathrm{s}}$ and $\mathbf{n}_{+\frac{1}{2}}$, we are able to induce, in controlled way, complex routes towards the final bipolar state. In the case of inhomogeneous shells, the specific symmetry of the shell allowed us to observe a hybrid splay-bend Helfrich wall for the first time.

\begin{description}
\item[PACS numbers]
May be entered using the \verb+\pacs{#1}+ command.
\end{description}
\end{abstract}

\pacs{Valid PACS appear here}
\maketitle


\section{\label{sec:level1}Introduction}

Topological defects are central to many areas of science, from particle physics to cosmology or materials engineering \cite{mermin1979topological, Kibble1976}.  They can be described from symmetry breaking considerations and, to a large extent, they determine the structure and physical properties of a material. Liquid crystals offer a unique playground for the study of topological defects because of the larger length-scales typically involved. In nematic liquid crystals, rod-like molecules exhibit long range orientational order, with the long axis of the molecules aligned along the director, defined by a unit vector $\bf n$ with head-tail symmetry ($\bf n = -\bf n$). This symmetry of the director enables nematic liquid crystals (NLCs) to accommodate a variety of topological defects, which can be easily produced and observed in the laboratory \cite{Kleman1989, Lavrentovich2001}. 

Besides their fundamental interest, topological defects in NLCs have also attracted attention for a wide range of applications \cite{Lavrentovich2001, Lagerwall2011rev}. The defect-mediated self-assembly of colloidal particles has emerged as a promising strategy to create nano/micro-structured materials with emergent new properties \cite{stark2001physics, Musevic2006col, lavrentovich2011liquid, Smalyukh2017}. Embedding a micro-sized particle in a uniform director field causes the disruption of the field, leading to the formation of topological defects in the vicinity of the particle \cite{poulin1997novel}. The anisotropic elastic interactions between defects associated with different particles induce the formation of colloidal structures, with a complexity that depends on the liquid crystal symmetry and the molecular anchoring at the particle surface. Additionally, nanoparticles can be trapped at the cores of topological defects, and patterned surfaces can therefore be engineered to introduce NLC defects as targeting sites for the assembly of colloidal particles into precisely controlled configurations \cite{Fleury2009, blanc2013ordering, Yoshida2015}.

\begin{figure}  
  \centering
  \includegraphics[width=8cm]{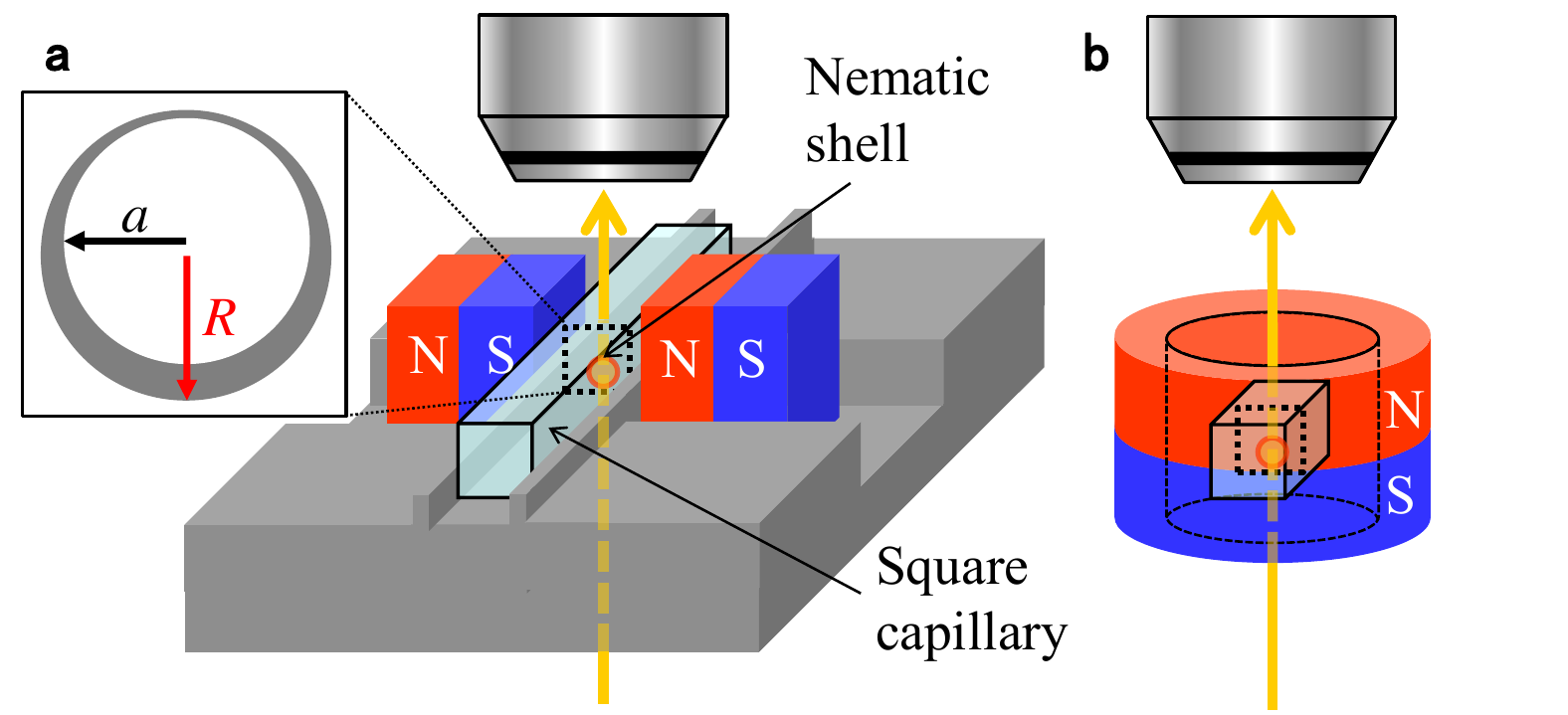}
  \caption{Experimental set-ups used to apply the magnetic field in a direction (a) perpendicular to gravity and (b) parallel to gravity.}
  \label{Fig1} 
  \end{figure}

Topological defects can also induce anisotropic colloidal interactions in a completely different way. Coating the spherical surface of a colloidal particle with a thin nematic shell induces the formation of an irreducible set of defects, as a result of geometrical frustrations in the orientational order of the liquid crystal \cite{nelson2002toward}. The ground state of very thin nematic shells has four defects sitting at the vertices of a tetrahedron. The coated sphere can then be viewed as a patchy colloidal particle with tetravalent coordination \cite{lubensky1992orientational, nelson2002toward, lopez2011frustrated}. The bonds between patches could be provided by chemical linkers attached at the four defects present in each colloid. The idea of using liquid crystals to produce colloids with a valence has fuelled the growth of research on liquid crystal shells \cite{Dzubiella2000, Vitelli2006, Bates2008, Bates2008b, Bates2009,  lopez2011frustrated, Kralj2011curv, liang2011nematic, Mirantsev2012geode, Napoli2012curv, Lopez-Leon2012a, Lopez-LeonSmect, Liang2012, Dhakal2012nema, Sec2012defe, Gharbi2013micro, Mondiot2013liqu, Liang2013, Seyednejad2013, Koning2013, Mbanga2014simu, Liang2014rigi, Lagerwall2015, Wand2015, Darmon2016a, Darmon2016, Mirantsev2016defe, Zhou2016, Koning2016, Tran2017, Mesarec2017curv, Sadati2017sphe, Qu2017trans, Urbanski2017Liq, Duan2017curv, Allahyarov2017smec, Nikoubashman2017semi, Khadilkar2018self, Serafin2018topo, Bates2018stab, Sharma2019real}. These shells have been produced by confining a nematic phase between two spherical aqueous interfaces, as schematically shown in Fig.\ref{Fig1}(a). The three-dimensional nature of these experimental shells enables a larger number of defect structures / particle valences. The predicted tetravalent defect structure has been observed in experimental shells, coexisting with a bivalent and a trivalent defect configurations \cite{fernandez2007novel}. Important efforts have been done in the last years to engineer shells with a controlled valence. Shell thickness and shape, molecular anchoring at the boundaries, or elasticity of the liquid crystal are just a few examples of parameters that have been studied theoretically, numerically and experimentally, with the goal of achieving such control \cite{Vitelli2006, Bates2008, Bates2008b, Bates2009, lopez2011frustrated, Kralj2011curv, Kralj2011curv, liang2011nematic,Lopez-Leon2012a, Lopez-LeonSmect, Liang2012, Dhakal2012nema, Sec2012defe, Mirantsev2012geode, Napoli2012curv, Liang2013,  Seyednejad2013, Koning2013, Lagerwall2015, Wand2015, Koning2016, Darmon2016a, Darmon2016, Mirantsev2016defe, Zhou2016, Tran2017, Mesarec2017curv, Sadati2017sphe, Bates2018stab, Sharma2019real}. Despite the progress, we are still far from having a robust approach to produce shells with a given defect structure.

The use of external fields has been proposed as a promising strategy to produce, in a controlled way, shells with a large spectrum of defect structures \cite{skavcej2008controlling, Oliveira2016}. Typical nematic liquid crystals have a larger dielectric constant (magnetic susceptibility) along the long molecular axis, and thus, the application of an electric (magnetic) field leads to the alignment of the director $\bf n$ along the direction of the field. Numerical studies have shown that the application of a strong uniform electric field causes structural changes in the tetrahedral configuration, which develops a bipolar structure with two surface defects or "boojums" at each spherical boundary. Remarkably, simulations reveal the formation of high-valence structures, such as an eight-defect structure, when quadrupolar fields are applied. Despite these promising predictions, the effect of external fields on liquid crystals shells has not been examined before in experiments.

In this work, we study the structural modifications undergone by tetravalent nematic shells under a strong and uniform magnetic field $\mathbf{B}$ using both experiments and simulations. The alignment of $\bf n$ with $\mathbf{B}$ triggers a series of structural transformations in the shells, which eventually adopt a bipolar configuration, as suggested by previous simulations. We uncover different scenarios in which the four $+1/2$ defects (inducing a $\pi$-rotation of $\bf n$) recombine by pairs to form two $+1$ boojums (inducing a $2\pi$-rotation of $\bf n$) on each shell boundary. The recombination process is mediated by the presence of inversion walls, which dynamically bring the defects together towards the poles. We show that the shape and nature of the walls depend on the relative orientation of the $+1/2$ defects with respect to the surface projection of the field $\mathbf{B}_{\mathrm{s}}$. 

 \begin{figure} [!ht]
  \centering
  \includegraphics[width=8.5cm]{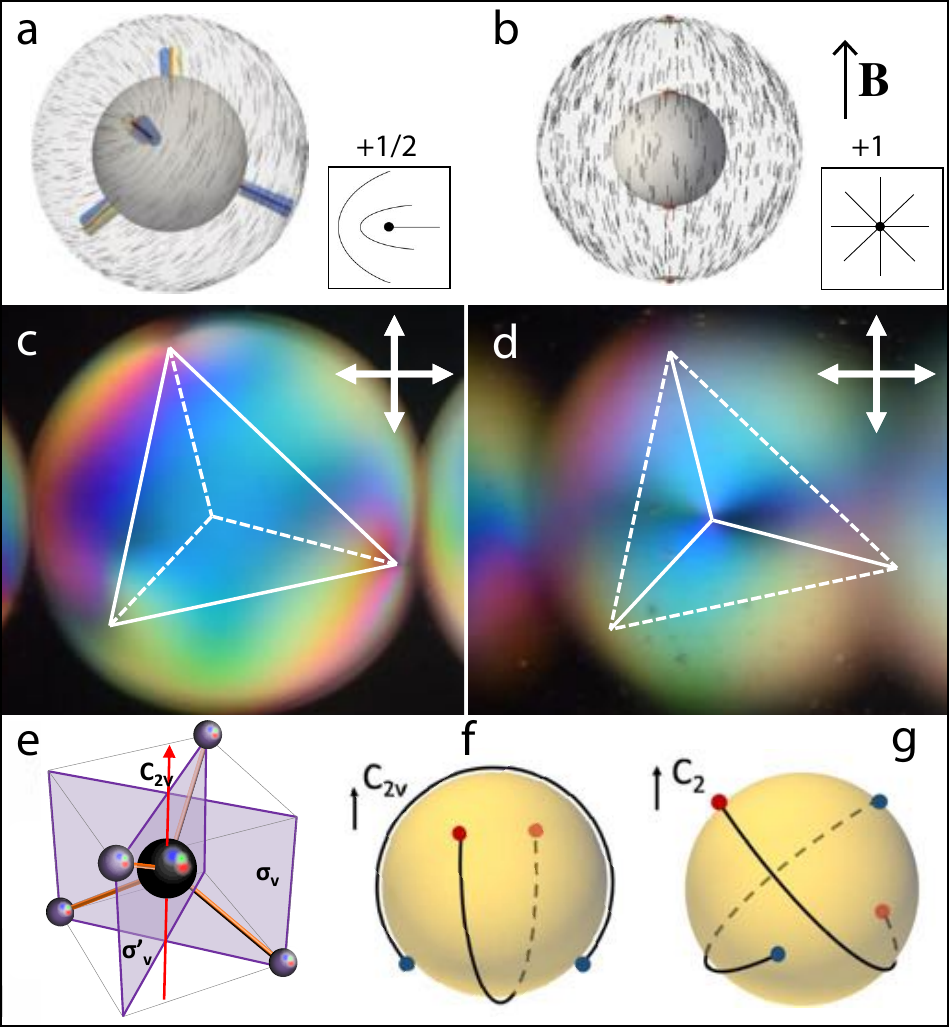}
  \caption{Tetrahedral defect structure in thin, homogeneous, nematic shells. (a) Simulated shell with tetrahedral defect structure. The shell has four $+1/2$ disclination lines arranged in a tetrahedral fashion. The inset shows a schematic representation of the director field on the surface of the shell, around each $+1/2$ defect. This structure is expected to evolve towards a bipolar defect structure, with a pair of $+1$ boojums at each pole, as shown in (b), when applying a sufficiently strong magnetic field. (c), (d) Experimental shell with tetrahedral defect structure. The two images are cross-polarized micrographies showing different focal planes of the same shell: three of the four $+1/2$ defects are in focus in (c), whereas the fourth $+1/2$ defect is in focus in (d). (e) Schematic illustration of one of the three $C_{2v}$ symmetry axes of a tetrahedron. (f) $C_{2v}$ axis and director stream line connecting the $+1/2$ defects in tetrahedral shells. (g) $C_{2}$ axis and director stream line connecting the $+1/2$ defects in tetrahedral shells.}
  \label{Fig2} 
  \end{figure}

 \begin{figure*} [!ht]
  \centering
  \includegraphics[width=13cm]{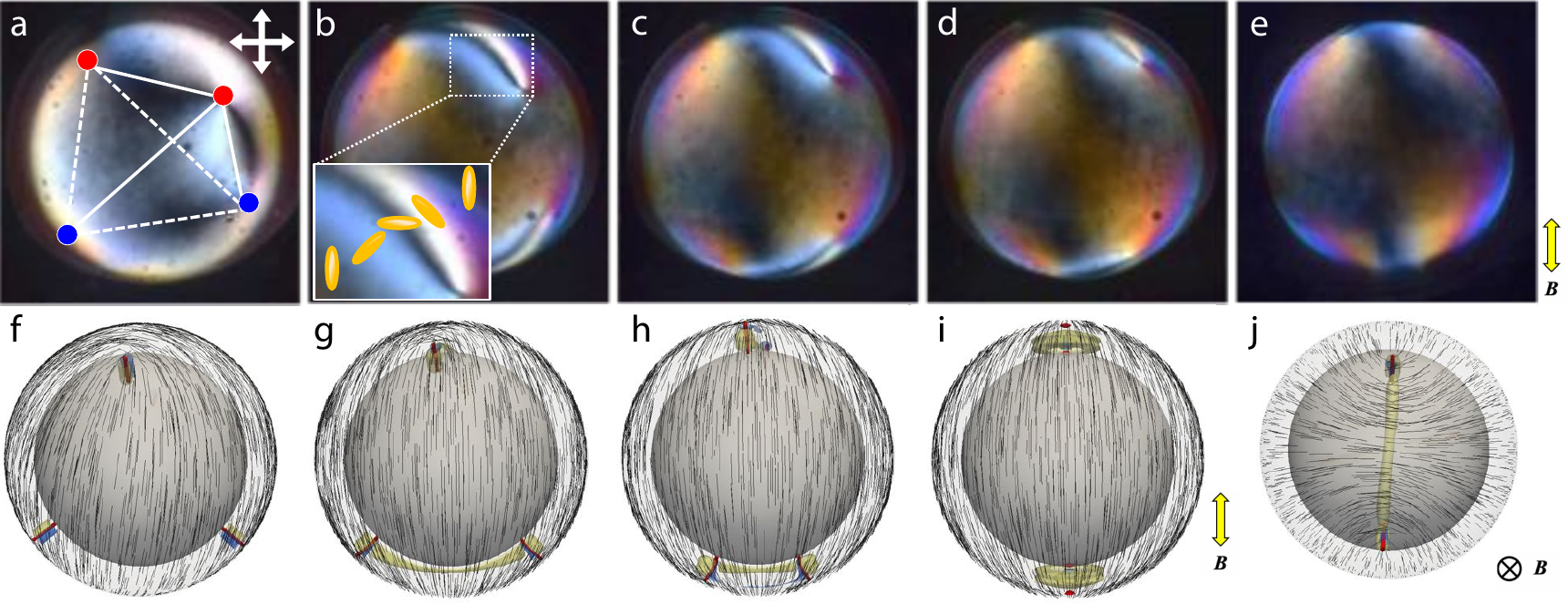}
  \caption{Structural transformations undergone by a nematic shell with tetrahedral defect structure upon applying a uniform magnetic field along the $C_{2v}$ axis of the shell. (a)-(e) Experimental cross-polarized images. (f)-(j) Simulated director field on the outer surface of the shell. (a), (f) Initial tetrahedral structure before applying the magnetic field. In the experimental images, defects of the same pair (connected by director streamlines) are represented with the same color. In the simulation plots, all the defects are represented in red (isosurface for $S=0.5$), while the splay and bend elastic distortions are represented in blue ($S_{\mathrm{SB}} > 0.005$) and yellow ($S_{\mathrm{SB}} < -0.005$), respectively. (b), (g) Formation of two inversion walls after applying the magnetic field. They run along geodesic lines, connecting defects of the same pair. The inset in (b) shows the change of birefringence near one of the walls, indicating a $\pi$-rotation of the director.  The way in which the director rotates across the walls, shown in (j), indicates that they are bend Helfrich walls. (c), (d) and (h) The walls are unstable and shrink over time while bringing the defects to the poles, eventually leading to the bipolar configuration shown in (e), (i).}
  \label{Fig3} 
\end{figure*}

\section{\label{sec:level1}Experiment details}

The experimental shells correspond to double emulsions produced in a conventional grass capillary microfluidic device \cite{utada2005monodisperse}. The middle phase is 4-n-pentyl-4’-cyanobiphenil (5CB), a liquid crystal that forms a nematic phase at room temperature. The inner and outer liquids are aqueous solutions containing 1wt\% of polyvinyl alcohol (PVA), which stabilizes the double emulsion and enforces planar anchoring of the liquid crystal at the inner and outer interfaces. The inset of Fig.\ref{Fig1}(a) shows a schematic representation of a shell. Typical values of the shell radius are  $R=50 - 100 \,\mu$m. The shell thickness, $h = R - a$, is on the order of several micrometers. Because of buoyancy effects and nematic elasticity, the experimental shells are inhomogeneous in thickness \cite{fernandez2007novel}. To obtain homogeneous shells, we produce extremely thin shells ($h < 1 \,\mu$m) by making the inner droplet bigger through osmotic swelling \cite{lopez2011frustrated}.

\begin{figure*} [!ht]
  \centering
  \includegraphics[width=13cm]{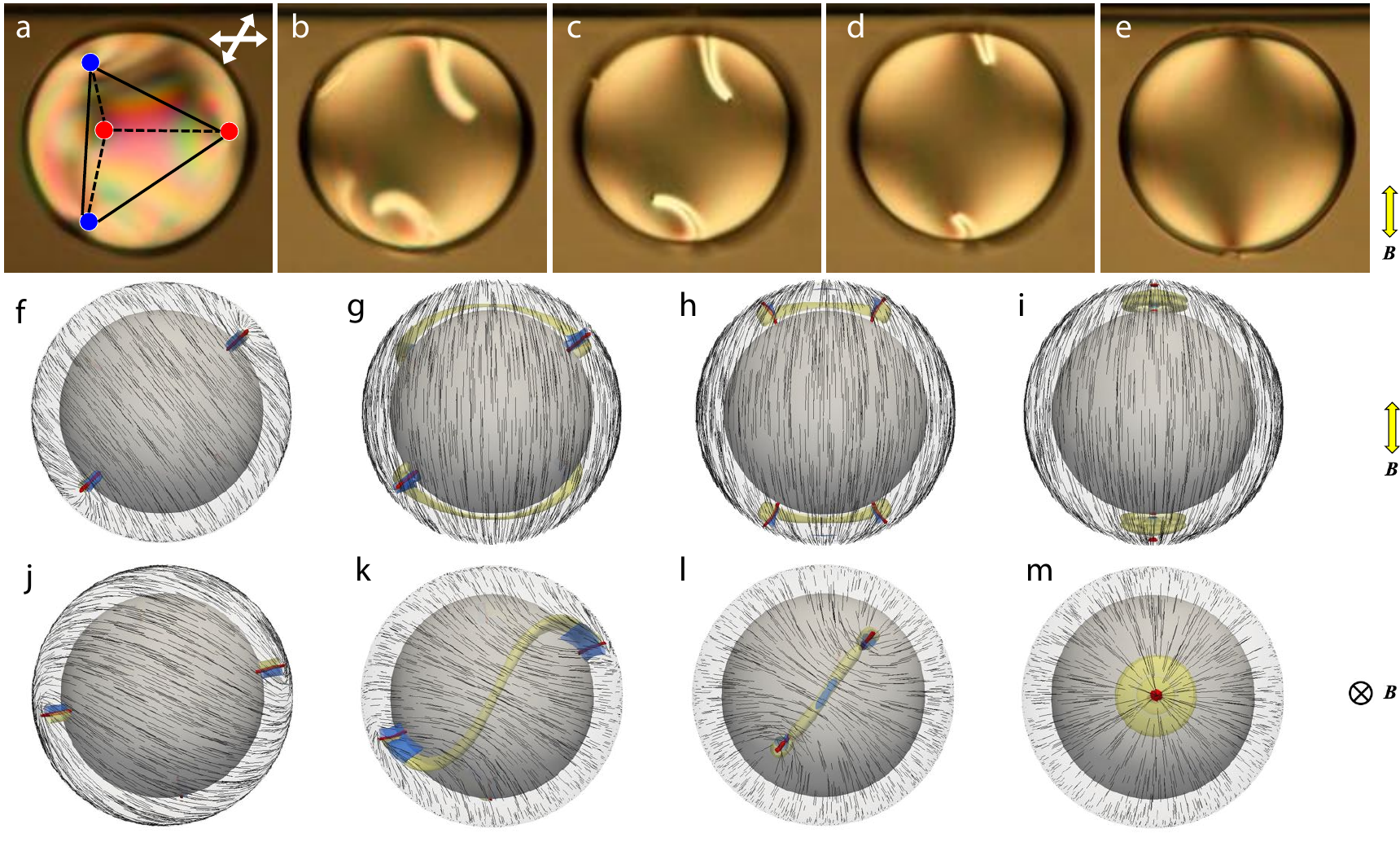}
  \caption{Structural transformations undergone by a nematic shell with tetrahedral defect structure upon applying a uniform magnetic field along the shell $C_{2}$ axis. (a)-(e) Experimental polarized images. (f)-(m) Simulated director field on the outer surface of the shell. (a), (f) and (j) Initial tetrahedral structure before applying the magnetic field. In the experimental images, defects of the same pair (connected by director streamlines) are represented with the same color. In the simulation plots, all the defects are represented in red (isosurface for $S=0.5$), while the splay and bend elastic distortions are represented in blue ($S_{\mathrm{SB}} > 0.005$) and yellow ($S_{\mathrm{SB}} < -0.005$), respectively. (b), (g) and (k) Formation of two inversion walls after applying the magnetic field. They run along curly paths (non-geodesic) and connect defects of different pair. The way in which the director rotates across the walls, shown in (k), indicates that they are bend Helfrich walls. (c), (d), (h) and (l) The walls are unstable and shrink over time. This makes the defects rotate and move to the poles, eventually leading to the bipolar configuration shown in (e), (i) and (m).}  
  \label{Fig4} 
\end{figure*}

After fabrication, the shells are collected in a $1$ mm inner diameter square capillary, which is placed in a sample holder made with a 3D printer. The capillary is then sandwiched between two permanent neodymium magnets, as schematically shown in Fig.\ref{Fig1}(a). This set-up permits application of strong ($ \approx$ 0.5 T) and uniform magnetic fields in a direction perpendicular to gravity. To apply magnetic fields parallel to the gravity direction, we place the sample inside a hollow cylindrical neodymium magnet, which also induces a strong ($ \approx$ 0.5 T), uniform field, see Fig.\ref{Fig1}(b). All experiments were performed at room temperature.

\section{\label{sec:level1}Simulation details}
We use a Landau-de Gennes continuum model for the order tensor $\bf Q$, which is defined by $Q_{ij} = S(n_i n_j - $ $\frac{1}{3} \delta_{ij})$ \cite{de1993physics}. The scalar order parameter is denoted by $S$. The total free energy is given by

\begin{align}
F({\bf Q})=& \int^{}_{\mathrm{bulk}} (\frac{A}{2}(1 - \frac{U}{3})Q_{ij}Q_{ji} - \frac{AU}{3}Q_{ij}Q_{jk}Q_{ki} \notag \\ 
& +\frac{AU}{4}(Q_{ij}Q_{ji})^2)dV \notag \\ 
& + \int^{}_{\mathrm{bulk}}\frac{L}{2} \frac{\delta Q_{ij}}{\delta x_k} \frac{\delta Q_{ij}}{\delta x_k}dV
 - \int^{}_{\mathrm{bulk}}\frac{1}{3} \epsilon_0 \epsilon_a ^{\mathrm{mol}} Q_{ij} E_i E_j dV \notag \\
& + \int^{}_{\mathrm{surf}}W (\tilde{Q}_{ij} - \tilde{Q}_{ij}^\perp)^2 d \Sigma
\end{align}

where $A$ is a material constant and $U$ is a dimensionless parameter that depends on temperature and pressure. A one-constant representation is adopted here, where $L$ denotes the elastic constant of the liquid crystal. The dielectric vacuum permittivity constant and molecular dielectric anisotropy are denoted by $\epsilon_0$ and $\epsilon_a ^{\mathrm{mol}} = (\epsilon_{//} - \epsilon_\perp) / S$, respectively. The anchoring strength is denoted by $W$. The order tensor ${\bf \tilde{Q}}$ is given by ${\bf Q} + S_{eq} {\bf I} /3$, where $S_{\mathrm{eq}}$ equals $\frac{1}{4}(1 + 3 \sqrt{1 - \frac{8}{3}U})$. The projection of ${\bf \tilde{Q}}$ onto the surface is denoted by ${\bf \tilde{Q}}^\perp = {\bf P \tilde{Q} P}$, where the projection operator ${\bf P}$ is defined by $P_{ij} = \delta_{ij} - v_i v_j$, and ${\bf v}$ is the unit vector normal to the surface \cite{fournier2005modeling}. The first term in Equation (1), which corresponds to enthalpic contributions to the free energy, serves to control the equilibrium value of the order parameter. The second term represents the elastic contributions to the free energy. It governs long-range director distortions and penalizes elastic deformations in the bulk \cite{doi2013soft}. The third term represents the energy due to the magnetic field \cite{Lozar2005wall}. The last term corresponds to the surface energy, which enforces degenerate planar anchoring on the shell surface. An iterative Ginzburg-Landau relaxation with finite differences on a cubic mesh (with resolution of 7.15 nm) is adopted to minimize the free energy \cite{ravnik2009landau}. To reveal the fine structure of defects, we use the splay-bend parameter $S_{\mathrm{SB}}$ constructed from second derivatives of the order parameter tensor ${\bf Q}$

\begin{equation}
S_{\mathrm{SB}} = \frac{\delta^2 Q_{ij}} {\delta x_i \delta x_j}.
\end{equation}

Large positive (negative) values of $S_{\mathrm{SB}}$ imply strong splay (bend) deformation. Polarization micrographs are calculated using the Jones $2 + 2$ matrix formalism, in which light travels along a chosen direction and the total phase shift is accumulated. The numerical parameters used in this work are $A = 1.067 \times 10^5 $ J/m$^3$, $U = 5, L = 6$ pN, $W = 1\times 10^{-3} $ J/m$^2$, $\hat{B} = 0.1$. The shells have an outer radius $R = 1 \,\mu$m. For homogenous shells, the inner radius is $a = 0.786 \,\mu$m. For inhomogeneous shells, the inner radius is $a = 0.643 \,\mu$m, and the shift between the centers of the inner and outer droplets is $d = 0.257 \,\mu$m.

\section{\label{sec:level1}Results and discussion}

Tetravalent shells exhibit four $+1/2$ defects, whose spatial distribution depends on the shell thickness gradient. In the next subsections, we study the structural transformations undergone by tetravalent shells in the presence of magnetic fields, stressing the role of the field direction with respect to the shell symmetries.

\subsection{\label{sec:level2}Homogeneous shell}

Because of the elastic repulsion between like-charged topological defects, in homogeneous shells, the four $+1/2$ defects are located at the vertices of a tetrahedron, see Fig.\ref{Fig2}(a) \cite{lubensky1992orientational, nelson2002toward, lopez2011frustrated}. The experimental realization of such a structure is shown in Fig.\ref{Fig2}(c) and (d), which are cross-polarized images of a thin, homogeneous nematic shell. Upon the application of a sufficiently strong magnetic field, $\mathbf{B}$, the tetrahedral structure evolves into a bipolar one, see Fig.\ref{Fig2}(b), as predicted by previous simulations \cite{skavcej2008controlling}, the bipolar axis being aligned with the magnetic field direction. Interestingly, we observe that the route towards the final bipolar state can be very different depending on the orientation of $\mathbf{B}$ with respect to the axes of symmetry of the tetrahedron. 

\begin{figure} [!ht]
  \centering
  \includegraphics[width=8cm]{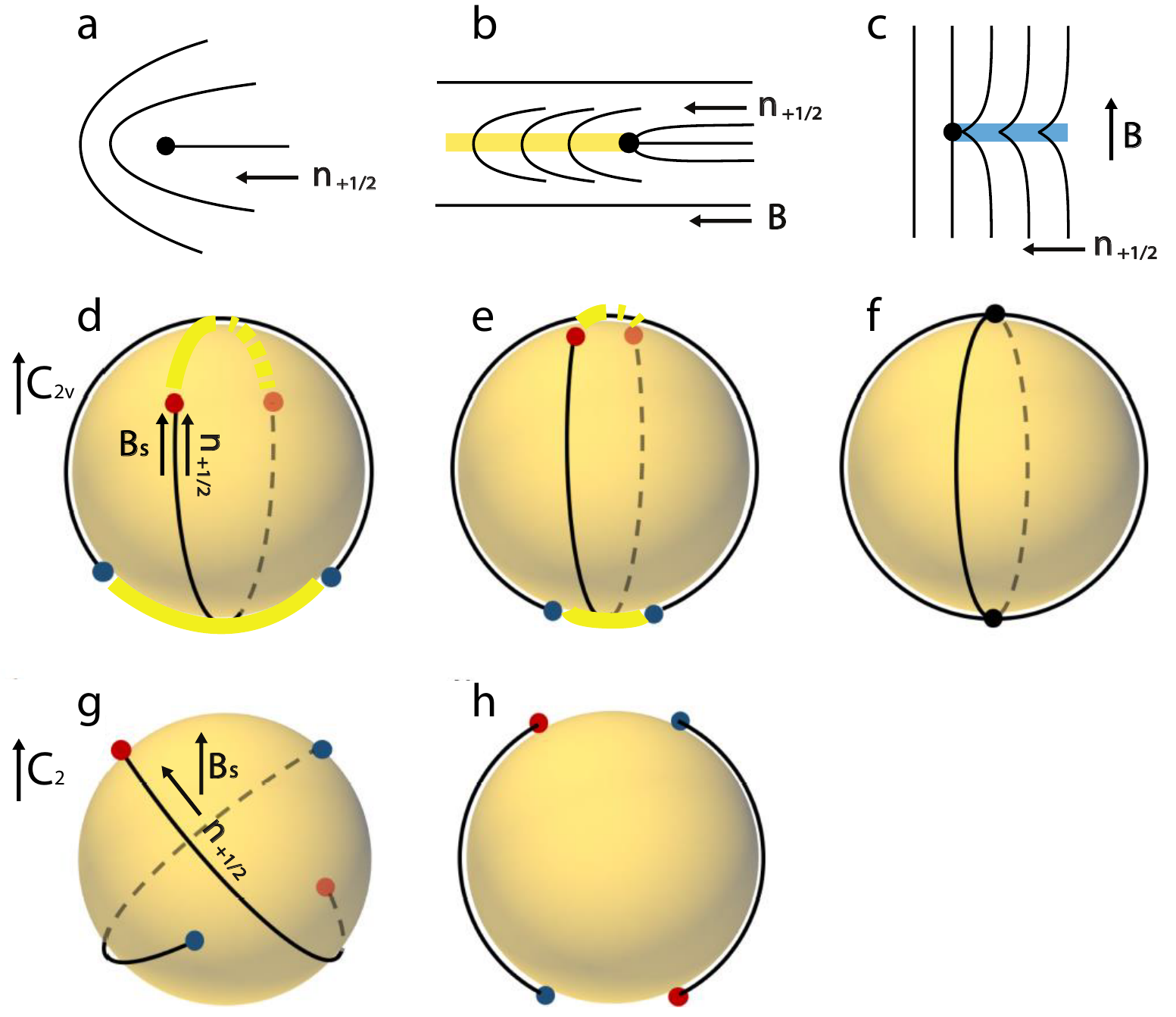}
  \caption{(a) Definition of the $\mathbf{n}_{+\frac{1}{2}}$ vector, denoting defect orientation. (b) When $\mathbf{B}$ is parallel to $\mathbf{n}_{+\frac{1}{2}}$, a bend-splay Helfrich wall (yellow stripe) appears along the $\mathbf{n}_{+\frac{1}{2}}$ direction. (c) When $\mathbf{B}$ is perpendicular to $\mathbf{n}_{+\frac{1}{2}}$, a splay-bend Helfrich wall (blue stripe) appears along the $-\mathbf{n}_{+\frac{1}{2}}$ direction. (d) In the case of a shell, when $\mathbf{B}$ is applied along the $C_{2v}$ axis, the surface projection of the field $\mathbf{B}_{\mathrm{s}}$ is parallel to $\mathbf{n}_{+\frac{1}{2}}$, and thus, two bend Helfrich walls appear between defects (yellow stripes). The great circle connecting the two defects of a pair is filled by a director streamline on one side (black line) and Helfrich wall on the other side (yellow stripe). (e), (f) Evolution of the walls and streamlines connecting defects with time. (g) When $\mathbf{B}$ is applied along the $C_{2}$ axis, $\mathbf{B}_{\mathrm{s}}$ and $\mathbf{n}_{+\frac{1}{2}}$ are not parallel, leading to the formation of Helfrich walls with a non-trivial shape (not represented). (h) These curvature walls connect defects from different pairs and shrink to eventually yield the bipolar configuration depicted in (f).} 
  \label{Fig5}
\end{figure}

A regular tetrahedron has three $C_{2v}$ symmetry axes: each $C_{2v}$ axis includes a two-fold rotational symmetric axis ($C_2$),  indicated by a red arrow in Fig.\ref{Fig2}(e), and two orthogonal mirror planes containing the $C_2$ axis, denoted by $\sigma_v$ and colored in purple in Fig.\ref{Fig2}(e). However, in a nematic shell, the director field around the defects, shown in Fig.\ref{Fig2}(a), breaks the degeneracy of the three $C_{2v}$ symmetry axes. To better illustrate this, in Figs.\ref{Fig2}(f) and (g), we have only represented the director streamlines connecting the $+1/2$ defects. Because of the symmetry of the streamlines, only one of the three $C_{2v}$ axes remains unaltered, see Fig.\ref{Fig2}(f), while the other two degrade to lower-order-symmetry axes, $C_2$, see  Fig.\ref{Fig2}(g). We apply a strong uniform magnetic field along the $C_{2v}$ and $C_2$ axes of homogeneous nematic shells and monitor the resulting defect motions. 

When $\mathbf{B}$ is applied approximately along the $C_{2v}$ axis, see Figs.\ref{Fig3}(a)-(e), two inversion walls suddenly appear, connecting the $+1/2$ defects by pairs. These walls produce a strong variation of the sample birefringence, which locally changes from dark to bright to dark at both sides of the wall, see the inset in Fig.\ref{Fig3}(b). This change of birefringence indicates a $\pi$-rotation of the director from one side to the other side of the wall. The $\pi$-walls, also called Helfrich walls, separate two regions of inverse alignment with the external field \cite{Chandrasekhar1992} and have been referred to as planar solitons \cite{Chandrasekhar1992}. With time, the $\pi$-walls become shorter bringing the two associated $+1/2$ defects to the poles, see Figs.\ref{Fig3}(c)-(e). The $+1/2$ defects approach each other by following the shorter path (geodesic): the $\pi$-walls run along great circles, where curvature is minimal. At the end of the process, which lasts several minutes, the shell adopts a bipolar configuration, with the bipolar axis aligned with the external field, see Fig.\ref{Fig3}(e). Depending on the strength of $\mathbf{B}$, the two $+1/2$ defects of each pair can either merge to form a $+1$ defect (strong fields) or stay close together (moderate fields). For the sake of simplicity, in the following, we will use the term \enquote{merging} for both cases.

\begin{figure} [!t]
   \centering
  \includegraphics[width=8cm]{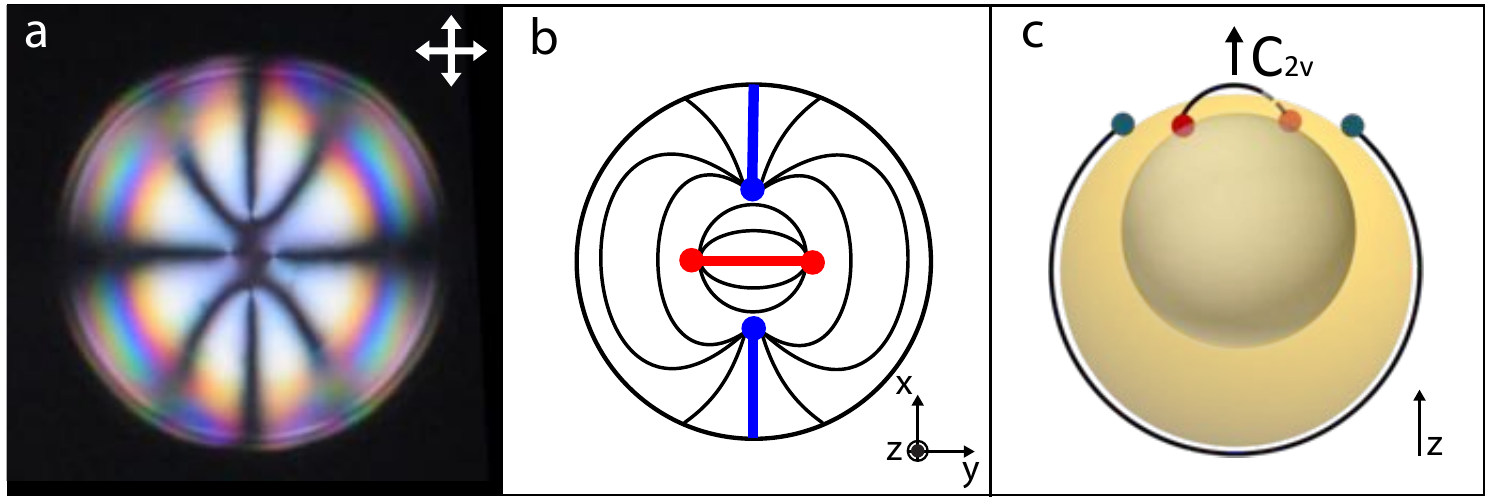}
 \caption{(a) Cross-polarized image of an inhomogeneous tetravalent shell. The four $+ 1/2$ disclinations are placed at the top of the shell, where it is thinner. (b) Schematic director field in an inhomogeneous tetravalent shell. (c) The blue defect pair is connected along the longest geodesic path, while the red defect pair is connected along the shortest geodesic path. This structure has just a $C_{2v}$ symmetry axis.}  
  \label{Fig6} 
\end{figure}

\begin{figure*} [!ht]
  \centering
  \includegraphics[width=14cm]{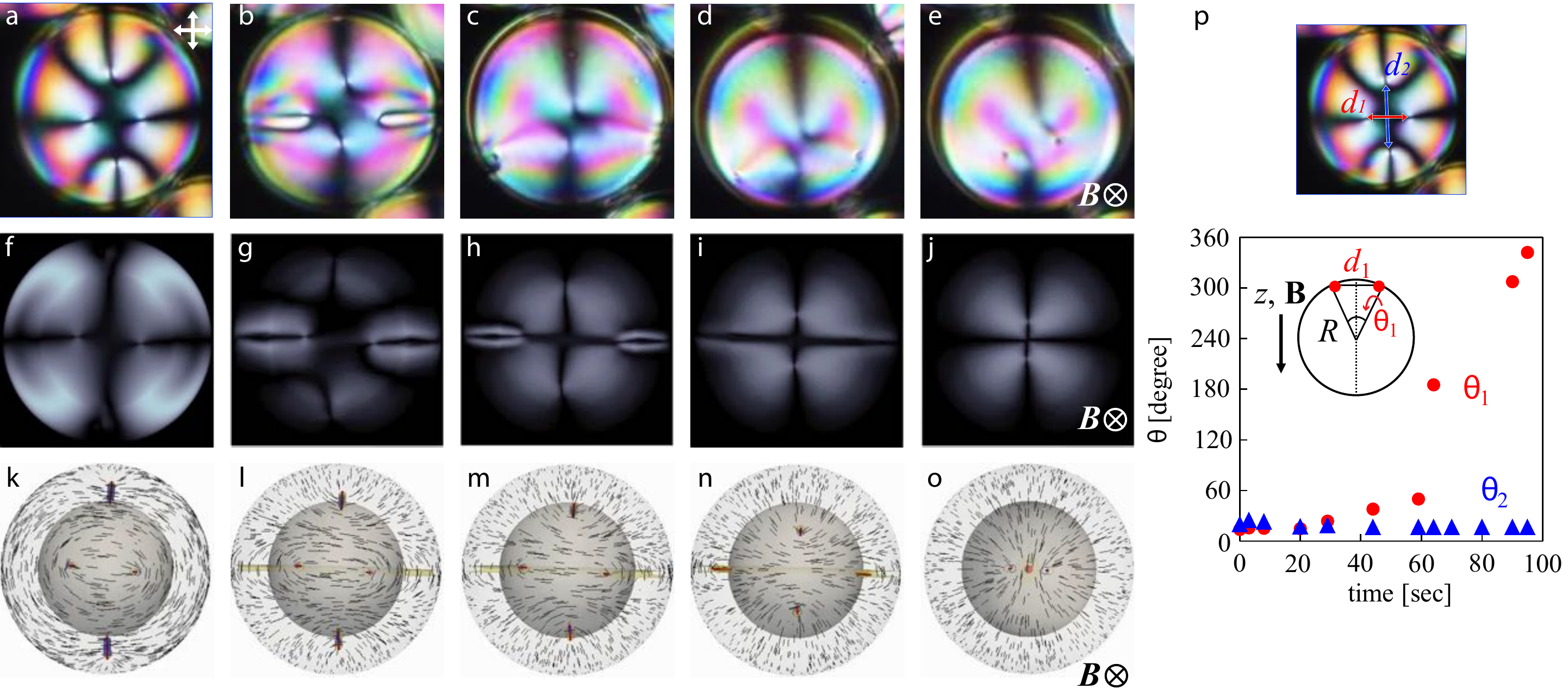}
  \caption{Structural transformations undergone by an inhomogeneous tetravalent nematic shell upon applying a uniform magnetic field along the $z$ axis. (a)-(e) Experimental cross-polarized images. (f)-(j) Simulated cross-polarized images. (k)-(o) Simulated director field on the outer surface of the shell. In the simulations, the defects are shown in red (isosurface for $S=0.5$), while the splay and bend elastic distortions are shown in blue ($S_{\mathrm{SB}} > 0.005$) and in yellow ($S_{\mathrm{SB}} < -0.005$), respectively. (a), (f) and (k) Initial state with the four defects placed at the top of the shell. (b), (g) and (l) Formation of two bend Helfrich walls after applying the filed, which run along geodesic paths, connecting defects of the same pair. (c), (h) and (m) The defects connected by the longest streamlines (shortest wall) get closer and coalesce at the top of the shell. (d), (i), (n)  The defects connected by the shortest streamlines (longest wall) move away from each other to eventually coalesce at the bottom of the shell.  (p) Time evolution of the angular distance  between defects ($\theta$) in experiments, showing that the two defect pairs behave asymmetrically.}  
  \label{Fig7} 
\end{figure*}

When $\mathbf{B}$ is applied approximately along the $C_2$ axis, see Figs.\ref{Fig4}(a)-(e), we observe an unexpected defect rotation. The two $\pi$-walls connecting the $+1/2$ defects by pairs do not run over geodesic lines but bend into wavy paths. The lines unwind while they shrink to eventually disappear when the defects reach the poles, see Fig.\ref{Fig4}(e).

Simulations allow us to extract additional information about the structural modifications in the shells under the effect of the magnetic field.  Before diving in, we take a glance at the possible structural transformations that can be induced by  $\mathbf{B}$ on single $+1/2$ defects, see Figs.\ref{Fig5}(a)-(c). Here, we denote the defect orientation as the direction pointed by the arrow $\mathbf{n}_{+\frac{1}{2}}$ in Fig.\ref{Fig5}(a). When  $\mathbf{B}$ is applied parallel to $\mathbf{n}_{+\frac{1}{2}}$, a bend-splay Helfrich wall appears in the $\mathbf{n}_{+\frac{1}{2}}$ direction, as shown in Fig.\ref{Fig5}(b). In this type of wall, the transition from $\mathbf{+n}$ to $\mathbf{-n}$ occurs mainly through a bend deformation (yellow stripe), although some splay is also present. When $\mathbf{B}$ is perpendicular to $\mathbf{n}_{+\frac{1}{2}}$, see Fig.\ref{Fig5}(c), it produces a splay-bend Helfrich wall instead, where the transition from $\mathbf{+n}$ to $\mathbf{-n}$ occurs mainly through a splay deformation (blue stripe). The wall appears again along the $-\mathbf{n}_{+\frac{1}{2}}$ direction, and thus, it is perpendicular to $\mathbf{B}$. For the sake of simplicity, here we will name these two types of inversion walls as \enquote{bend wall} and \enquote{splay wall}, respectively.

When the external field $\mathbf{B}$ is applied along the $C_{2v}$ axis, its projection on the surface ($\mathbf{B}_{\mathrm{s}}$) at the defect position is parallel to $\mathbf{n}_{+\frac{1}{2}}$, see Fig.\ref{Fig5}(d). In this situation, we expect to see the formation of bend walls between $+1/2$ defects, represented as a yellow stripe in Figs.\ref{Fig5}(d) and (e). This is confirmed by our simulations: Figs.\ref{Fig3}(g) and (j) show one of these inversion walls with major bend distortions in yellow ($S_{\mathrm{SB}} < -0.005$). Since the wall bears high elastic energy, it shrinks and makes the defects approach until they merge into a $+1$ defect, see Fig.\ref{Fig3}(i). The complete transformation is shown in Figs.\ref{Fig3}(f)-(i) and Figs.\ref{Fig5}(d)-(f), where Fig.\ref{Fig3}(i) and Fig.\ref{Fig5}(f) illustrate the final bipolar configuration. 

Our simulations also explain the origin of the defect rotation occurring when the external field is applied parallel to the $C_2$ axis. In this case, $\mathbf{n}_{+\frac{1}{2}}$ and $\mathbf{B}_{\mathrm{s}}$ are no longer parallel to each other, as shown Fig.\ref{Fig5}(g). Near the defects, the inversion walls follow $\mathbf{n}_{+\frac{1}{2}}$, but continuously wind themselves to be aligned with $\mathbf{B}_{\mathrm{s}}$ in the region between the two defects. That leads to the formation of s-shape walls, see Figs.\ref{Fig4}(g) and (k), which connect defects of different pairs. The values of the bend-splay parameter $S_{\mathrm{SB}}$ indicate that the formed wall is a bend wall. As the wall shrinks, the defects are pulled along, see Figs.\ref{Fig4}(h) and (l), and eventually align their orientation $\mathbf{n}_{+\frac{1}{2}}$ with $\mathbf{B}_{\mathrm{s}}$. The two $+1/2$ defects reach the pole and fuse together when the wall disappears, see Figs.\ref{Fig4}(i) and (m).

\subsection{\label{sec:level2}Inhomogeneous shell}

\begin{figure*} [!ht]
  \centering
  \includegraphics[width=15cm]{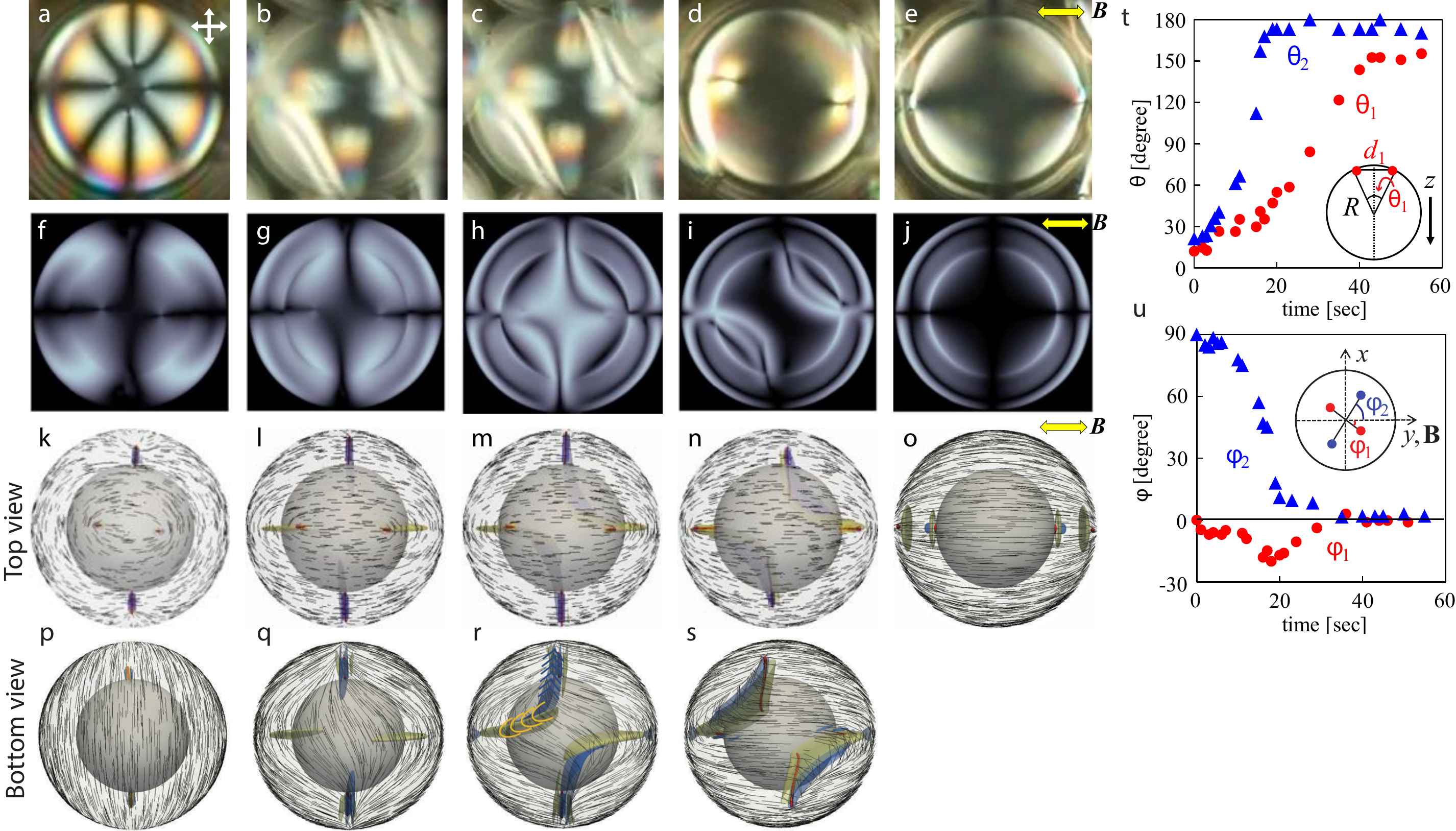}
  \caption{Structural transformations undergone by an inhomogeneous tetravalent nematic shell upon applying a uniform magnetic field along the $y$ axis. (a)-(e) Experimental cross-polarized images. (f)-(j) Simulated cross-polarized images. (k)-(s) Simulated director field on the outer shell surface: (k)-(o) top view, (p)-(s) bottom view. In the simulations, the defects are shown in red (isosurface for $S=0.5$), while the splay and bend elastic distortions are shown in blue ($S_{\mathrm{SB}} > 0.005$) and in yellow ($S_{\mathrm{SB}} < -0.005$), respectively. (a), (f), (k) and (p) Initial state with the four defects placed at the top of the shell. (b), (g), (l) and (q) Formation of two parabolic Helfrich walls after applying the field, connecting defects of different pairs. (c), (h),(m) and (r) The hyperbolic walls have a hybrid nature: as highlighted in (r), they stem from the junction of a bend Helfrich wall with a splay Helfrich wall. (d), (i), (n) and (s) The walls are unstable and shrink while pulling the associated defects towards opposite poles in the $\mathbf{B}$ direction.  (e), (j) and (o) Final bipolar configuration. (t) Time evolution of the angular distance between defects ($\theta$) in experiments, showing that the two defect pairs behave asymmetrically. (u) Defect rotation in the xy plane ($\varphi$) as a function of time in experiments. The red pair of defects rotates by $\pi/2$ to get aligned with the blue pair.}  
  \label{Fig8} 
\end{figure*}

Due to buoyancy and nematic elasticity, nematic LC shells are usually heterogeneous in thickness. In this geometry, the four $+1/2$ defects are located in the thinner part of the shell to reduce the bulk elastic energy \cite{fernandez2007novel}, as shown in Fig.\ref{Fig6}(a), which is a cross-polarized image of a heterogeneous shell. In this geometry, the director field connects the $+1/2$ defects in two asymmetric pairs, as schematically represented in Figs.\ref{Fig6}(b) and (c): one pair is connected along the longest geodesic path (blue pair), while the other one is connected along the shortest geodesic path (red pair). This structure is less symmetric than the tetrahedral one: the $C_{2v}$ axis remains, as indicated in Fig.\ref{Fig6}(c), but the two $C_2$ axes no longer exist. Here, we denote the $C_{2v}$ axis as the $z$ axis and the directions connecting the blue and red defect pairs as the $x$ and $y$ axes, respectively, see  Fig.\ref{Fig6}(b). 

We first examine the effect of $\mathbf{B}$ when it is applied along the $z$ axis ($C_{2v}$ axis). In this case, $\mathbf{B}_{\mathrm{s}}$ is again roughly parallel to the orientations of the four defects, see Fig.\ref{Fig6}(b). The transformation, shown in Figs.\ref{Fig7}(a)-(e) for experiments and Figs.\ref{Fig7}(f)-(o) for simulations, is very similar to that in homogeneous shells, except that the two bend inversion walls are not equal in length. As a result, the two pairs of defects move asymmetrically: in the blue pair, the defects get closer before merging together, while in the red pair, they move away from each other to reach the opposite hemisphere, where they eventually merge. This asymmetric behavior becomes evident when plotting the evolution of the angular distance between defects in each pair, $\theta_1$ (red pair) and $\theta_2$ (blue pair), as a function of time, see Fig.\ref{Fig7}(p).  

\begin{figure*} 
  \centering
  \includegraphics[width=14cm]{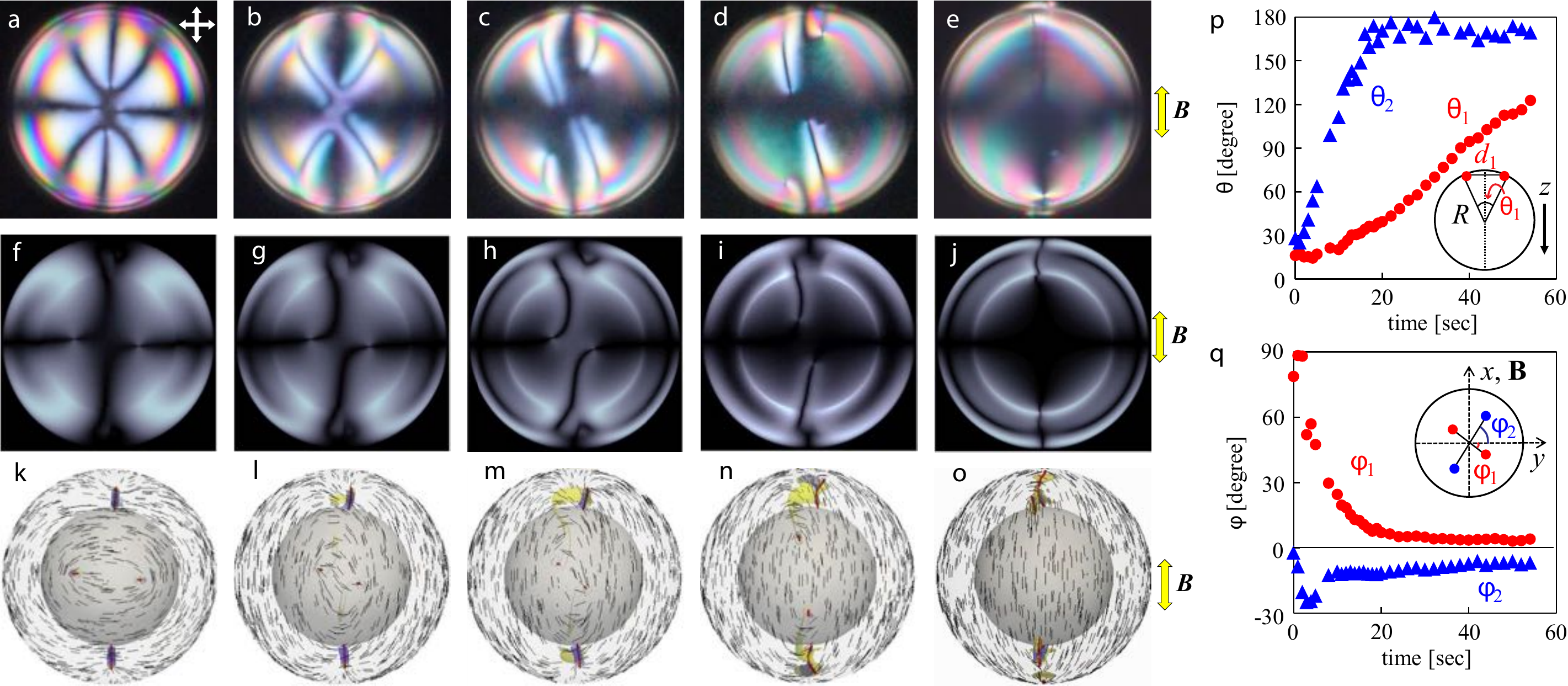}
  \caption{Structural transformations undergone by an inhomogeneous tetravalent nematic shell upon applying a uniform magnetic field along the $x$ axis. (a)-(e) Experimental cross-polarized images. (f)-(j) Simulated cross-polarized images. (k)-(o) Simulated director field on the outer shell surface. In the simulations, the defects are shown in red (isosurface for $S=0.5$), while the splay and bend elastic distortions are shown in blue ($S_{\mathrm{SB}} > 0.005$) and in yellow ($S_{\mathrm{SB}} < -0.005$), respectively. (a), (f) and (k) Initial state with the four defects placed at the top of the shell. (b), (g) and (l) Formation of two bend Helfrich walls after applying the field, connecting defects of different pairs. (c), (h) and (m) The initial open-loop shape of the walls evolves into a u-shape with time. (d), (i), (n) and (s) The walls are unstable and shrink while bringing the associated defects towards opposite poles in the $\mathbf{B}$ direction.  (e), (j) and (o) Final bipolar configuration. (p) Time evolution of the angular distance between defects ($\theta$) in experiments, showing that the two defect pairs behave asymmetrically. (q) Defect rotation in the xy plane ($\varphi$) as a function of time in experiments. The red pair of defects rotates by $\pi/2$ to get aligned with the blue pair.} 
  \label{Fig9} 
\end{figure*}

\begin{figure} 
  \centering
  \includegraphics[width=7cm]{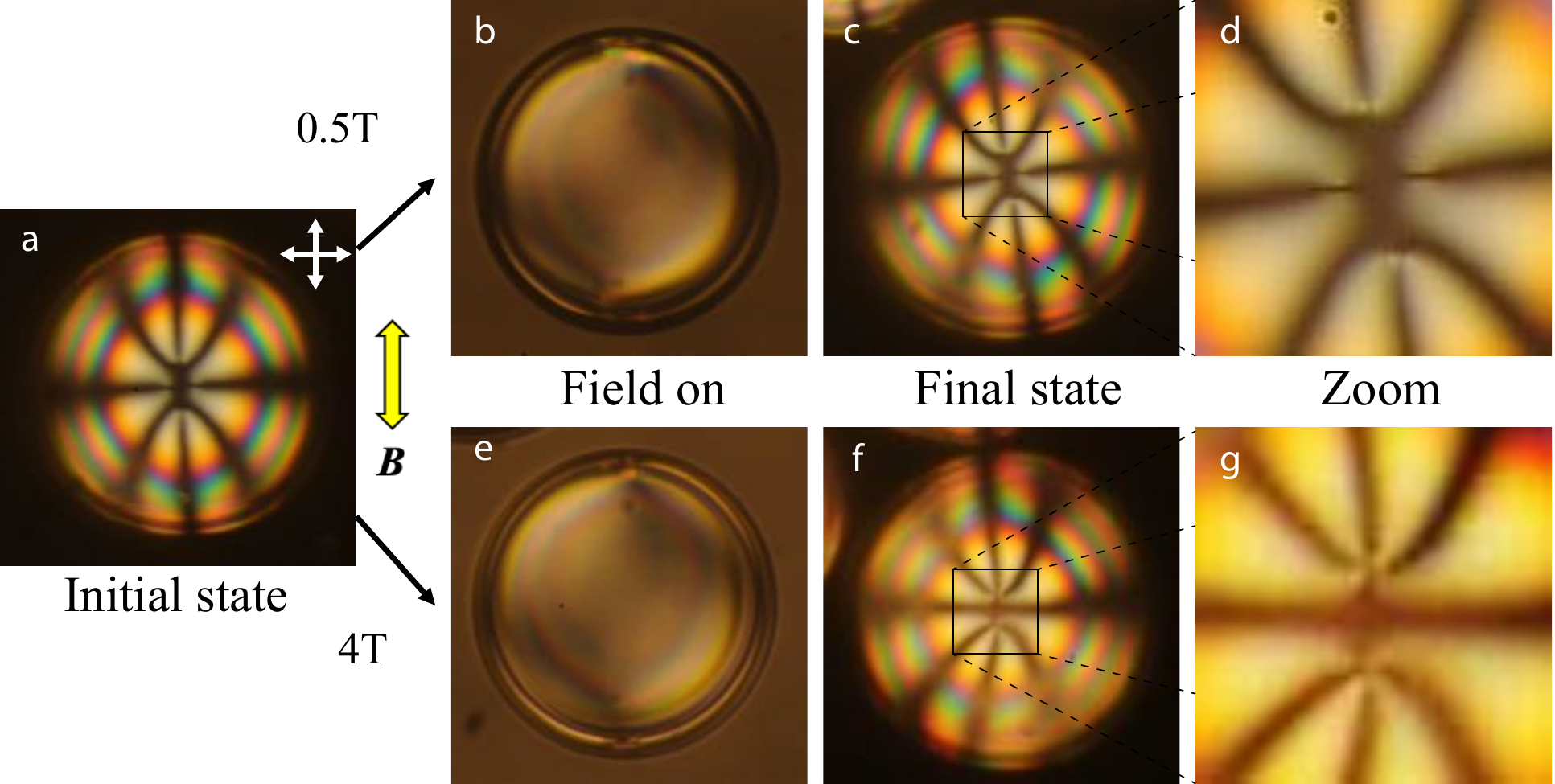}
  \caption{Effect of the magnetic field strength. (a) Initial state with four $1+/2$ disclinations placed at the top o the shell. (b) Applying a 0.5T magnetic field makes the shell adopt a bipolar structure, with a pair of close $+1/2$ disclinations at each pole. (c), (d) When the field is off, the structure relaxes back to the initial state. This process is reversible. (e) Applying a 4T magnetic field also makes the shell adopt a bipolar structure, but in this case the two $+1/2$ disclinations at the poles fuse together and give rise to two $+1$ boojums, one located above the other on the inner and outer surfaces of the shell. (f), (g) This transformation is irreversible: when the field is off, the boojum defect structure remains.}   \label{Fig10}
\end{figure}

When the magnetic field $\mathbf{B}$ is parallel to the $y$ axis, the defect orientation of the red pair is parallel to $\mathbf{B}_s$ while that of the blue pair is perpendicular to $\mathbf{B}_{\mathrm{s}}$, see Fig.\ref{Fig6}(b). According to our previous discussion about single $+1/2$ defects on the plane (Figs.\ref{Fig5}(a)-(c)), we expect to observe the emergence of a bend wall along the $\mathbf{n}_{+\frac{1}{2}}$ direction of the red pair, and a splay wall along the $-\mathbf{n}_{+\frac{1}{2}}$ direction of the blue pair. The simulated evolution of the director field and the defect evolution are shown in Figs.\ref{Fig8}(k)-(o). Upon application of the magnetic field, a bend wall and splay wall nucleate and propagate from different defect pairs as predicted, Figs.\ref{Fig8}(l) and (q), until they join each other and form a hybrid inversion wall. This wall has a hyperbolic shape, see Figs.\ref{Fig8}(m) and (r), with the bend part mostly parallel to $\mathbf{B}_{\mathrm{s}}$ and splay part mostly perpendicular to $\mathbf{B}_{\mathrm{s}}$. The hybrid walls can also be visualized in the simulated polarization micrographs shown in Figs.\ref{Fig8}(h),(i).

Experimentally, when we apply the magnetic field along the $y$ direction, we observe a solid rotation of the shell around the $C_{2v}$ axis. This rotation brings the shell to the situation studied in Fig.\ref{Fig9}, where the director in the thick hemisphere of the shell is aligned with the magnetic field. We can avoid this phenomenon by bringing the shell into contact with the glass walls of the observation capillary, where friction forces prevent the shell from rotating. Despite the images being blurry, this trick allows us to study this interesting configuration. We observe the formation of two hyperbolic walls connecting defects of different pairs, in agreement with the simulation results. To our knowledge, such a hybrid inversion wall has not been reported before, since it requires curved substrates and non-trivial defect configurations, as those present in heterogeneous tetravalent shells.  After the formation of the hybrid inversion wall, the defects move along the wall to eventually merge together (Fig.\ref{Fig8}(e)). Here again the behavior of the two pairs of defects is asymmetric: the blue pair recombines before the red one, see the temporal evolution of $\theta_1$ and $\theta_2$ in Fig.\ref{Fig8}(t). Besides, the pair connected by the longest geodesic path (blue pair) undergoes a $\pi/2$ rotation with respect to the y axis, while the pair connected by the shorter geodesic path (red pair) keeps its initial orientation, see the evolution of $\varphi_1$ and $\varphi_2$ in Fig.\ref{Fig8}(u). This operation aligns the defect orientations of the two defect pairs.

When $\mathbf{B}$ is applied along the $x$ axis, the general rule for single defects fails: it predicts a splay wall connecting the red defect pair, which is not observed either in experiments or simulations. This is probably due to the crowding of defects in the thinner part of shell. Instead, the bend wall arising from the blue pair winds back and attaches itself to the red pair, as shown in Fig.\ref{Fig9}(c). Here again, the walls connect defects from different pairs. The walls have an open-loop shape that evolves into a u-shape as the walls shrink, see Figs.\ref{Fig9}(b)-(d). Figs.\ref{Fig9}(p) and (q) show the evolution of each defect pair in terms of $\theta$ and $\varphi$ during the transition. The blue pair moves along the $\mathbf{B}_{\mathrm{s}}$ direction to the pole directly. While the red defects first approach towards each other, and then move apart when the curved bend wall forms. The red pair keeps rotating until it becomes parallel to the blue pair, while the distance between defects progressively shortens. The experimental observations are in good agreement with our simulation calculations, shown in Figs.\ref{Fig9}(f)-(o).

Finally, we would like to note that while the presence of a magnetic field induces structural re-arrangements that eventually lead to the formation of a bipolar structure, the $+1$ defects at the poles can be either $+1$ boojums or pairs of close $+1/2$ disclinations. Although both situations yield the same far field, the structures are topologically non-equivalent. The intensity of the magnetic field dictates whether the $+1/2$ disclinations completely fuse together to give a $+1$ boojum or not . Fig.\ref{Fig10} shows how imposing a 0.5T magnetic field on a shell that initially has four $+1/2$ disclinations makes it adopt a bipolar structure, which relaxes towards the initial state when the field is off, indicating that the $+1/2$ disclinations did not fuse together. Conversely, applying a $4 T$ magnetic field leads to the irreversible formation of $+1$ boojums that relocate in the thin part of the shell when the field is off, demonstrating the ability of magnetic fields to change the valence of the shell.

\section{Conclusion}

In this work, we study the structural transformations that arise in tetravalent nematic shells under a magnetic field, using both experiments and simulations. The presence of a sufficiently strong magnetic field makes the $+1/2$ defects move in pairs towards the poles of the shell. The defect trajectories and dynamics are controlled by inversion walls, which appear when the field is applied, connecting the $+1/2$ defects by pairs. These inversion walls are unstable and shrink over time, pulling the two $+1/2$ defects attached to their ends together. Depending on the strength of the field, the two defects either fuse together (strong fields) or just stay close by (moderate fields) when they reach the poles. At the end of the transformation, the shells adopt a bipolar configuration, with the bipolar axis aligned along the field. The nature and shape of the inversion walls, and thus the defect trajectories and dynamics, depend on the relative orientation between $\mathbf{B}_{\mathrm{s}}$, the local projection of the magnetic field on the shell surface, and $\mathbf{n}_{+\frac{1}{2}}$, a vector describing the orientation of the defects.  By analyzing the motion of the defects in homogeneous shells, we observed that, when $\mathbf{B}_{\mathrm{s}}$ is parallel to $\mathbf{n}_{+\frac{1}{2}}$, straight bend Helfrich walls emerge between defect pairs. This situation arises when the field is applied along the $C_{2v}$ symmetry axis of the shell. In contrast, when there is some angle between $\mathbf{B}_{\mathrm{s}}$ and $\mathbf{n}_{+\frac{1}{2}}$, curved Helfrich bend walls form, inducing rotation in the defect motion. This situation occurs when the field is applied along the $C_{2}$ symmetry axis of the shell. This general behavior is reproduced in inhomogeneous tetravalent shells, whose specific symmetry allowed us to observe hybrid splay-bend Helfrich walls for the first time. Our experimental observations are confirmed by numerical results, validating the role of the inversion walls in the mechanism of defect reorganisation under the effect of external fields. A better understanding of this mechanism may provide new methods to not only control the defect position, number and valence, but also to manipulate defect dynamics. 

\bibliography{References}
\bibliographystyle{rsc} 

\end{document}